\documentclass[aps,12pt,prb,preprint,superscriptaddress,longbibliography]{revtex4-1}
\usepackage{color}
\usepackage{fancyhdr} 
\usepackage{varwidth} 
\usepackage{graphicx,subfigure}  
\usepackage{bm} 
\usepackage{listings}
\usepackage{enumerate}%
\usepackage[colorlinks,citecolor=blue,linkcolor=blue]{hyperref}
\usepackage{amsmath}
\usepackage{threeparttable}
\usepackage{graphicx}
\usepackage{amssymb}
\usepackage{amsthm}
\usepackage{rotating}
\usepackage{multirow}

\begin{document}

\title{Coexistence of multiple interfacial states at heterogeneous solid/liquid interface}
\author{Jiaojiao Liu}
\affiliation{State Key Laboratory of Precision Spectroscopy, School of Physics and Electronic Science, East China Normal University, Shanghai 200241, China}
\author{Hongtao Liang}
\affiliation{State Key Laboratory of Precision Spectroscopy, School of Physics and Electronic Science, East China Normal University, Shanghai 200241, China}
\affiliation{Research and Development Department, Zhangjiang Laboratory, Shanghai 201204, China}
\author{Jinfu Li}
\affiliation{School of Materials Science and Engineering, Shanghai Jiao Tong University, Shanghai 200240, China}
\author{Brian B. Laird}
\affiliation{Department of Chemistry, University of Kansas, Lawrence, Kansas 66045, United States}
\author{Yang Yang}
\affiliation{State Key Laboratory of Precision Spectroscopy, School of Physics and Electronic Science, East China Normal University, Shanghai 200241, China}
\date{\today}

\begin{abstract}
The growing trend towards engineering interfacial complexion (or phase) transitions has been seen in the grain boundary and solid surface systems. Meanwhile, little attention has been paid to the chemically heterogeneous solid/liquid interfaces. In this work, novel in-plane multi-interfacial states coexist within the Cu(111)/Pb(l) interface at a temperature just above the Pb freezing point is uncovered using atomistic simulations. Four monolayer interfacial states, i.e., two CuPb alloy liquids and two pre-freezing Pb solids, are observed coexisting within two interfacial layers sandwiched between the bulk solid Cu and bulk liquid Pb. Through computing the spatial variations of various properties along the direction normal to the in-plane solid-liquid boundary lines for both interfacial layers, a rich and varied picture depicting the inhomogeneity and anisotropy in the mechanical, thermodynamical, and dynamical properties is presented. The ``bulk'' values extracted from the in-plane profiles suggest that each interfacial state examined has distinct equilibrium values from each other and significantly deviates from those of the bulk solid and liquid phases, and indicate that the ``complexion (or phase) diagrams'' for the Cu(111)/Pb(l) interface bears a resemblance to that of the eutectic binary alloy systems, instead of the monotectic phase diagram for the bulk CuPb alloy. The reported data could support the development of interfacial complexion (or phase) diagrams and interfacial phase rules and provide a new guide for regulating heterogeneous nucleation and wetting processes.
\end{abstract}

\maketitle

\section{Introduction}
\label{sec:Intro}

Since the term interfacial ``complexion'' -- thermodynamic equilibrium states of an interface -- having been introduced and recognized\cite{Tang06a} over the past two decades, see in Ref.\onlinecite{Cantwell14} and references therein, numerous grain boundary complexions with different segregation feature and thicknesses in ceramics and alloys have been unveiled. The manipulation of the interfacial complexion transition offers potential engineering of various interfacial-related processes (e.g., sintering\cite{Gupta07}, embrittlement\cite{Luo11}, and crystal growth\cite{Dillon07}) and interfacial properties (e.g., interfacial mobility\cite{Holm03}, atomic or ionic diffusivities\cite{Divinski12,Bowman17}, electrical or thermal conductivities\cite{Marvel18,Zong17}.

The majority of recent studies of the multiple interfacial complexion (or phase) coexistence and transition are made in grain boundary systems\cite{Heckman18,Brien18,Peter18,Meiners20,Barr21,Tuchinda22} and solid surfaces\cite{Wynblatt07,Pickering18,Tang21,Wang21}. In contrast, the idea that the chemically heterogeneous solid/liquid interfaces can also exhibit multiple interfacial complexion (or phase) coexistence and transitions, similar to the grain boundary and surface systems, is rare. Such disparity is because the direct experimental characterization on the interfacial layers for the solid/liquid interfaces is more challenging than for the free surfaces and grain boundaries\cite{Gabrisch01,Oh05,Moorthy13,Schneider17}. With the aid of advanced computer simulation techniques, several novel interfacial structural states and interfacial phase transitions are uncovered, such as the lateral and layer ordering of the interfacial liquids\cite{Yang14,Kern20}, novel pre-freezing\cite{Palafox16}, roughening and premelting transitions\cite{Yang13}. Few existing study\cite{Liang18} suggests two interfacial states (or phases) can in-plane coexist within the heterogeneous solid/liquid interface.

Compared with the experimental and computational achievements, the theoretical frameworks regarding the statistical mechanical nature of the in-plane interfacial complexion (or phase) transitions remain less developed. For example, several types of phenomenological models have been proposed to qualitatively explain the various transitions among interfacial complexions (or phases)  (lattice gas-based model\cite{Wynblatt05,Luo09,Rickman16}, regular solution-based model\cite{Wynblatt08}, diffuse-interface model\cite{Tang06b}, and force-balance model\cite{Clarke93}), in addition, a theoretical attempt by Frolov and Mishin has been proposed to extend thermodynamic formalism of the phase equilibria and phase rules into semi-2D interfacial systems\cite{Frolov15b}. Meanwhile, efforts have been devoted to developing interfacial complexion (or phase) diagrams\cite{Mishin09,Luo12,Gao19}, see more details in Ref.\onlinecite{Cantwell14}. However, most of these efforts focused on the simplified version of the interfacial complexion (or phase) diagram, which could not be verified by experiment yet. Therefore, existing thermodynamical models in predicting interfacial complexion (or phase) diagrams and transitions have not been rigorously validated, thus hindering the breakthrough regarding the precise control of the interfacial structure and various properties affected by the interfacial complexion (or phase) transitions.

In this study, based on the equilibrium molecular-dynamics (MD) methods and the embedded atom model, we report an atomistic simulation of a chemically heterogeneous Cu(111)/Pb(liquid) interface, in which four interfacial liquid and interfacial solid states, two boundary lines, coexist within two atomic layers sandwiched between the bulk solid Cu and liquid Pb. Note that the term ``interfacial state'' will be used more frequently in describing the current simulation results, rather than ``interfacial complexion (or phase)'', to avoid potential terminological disagreements. The elegant solid/liquid in-plane equilibria among these interfacial states and their thermodynamics and mechanical properties are carefully characterized by computing the in-plane profiles of the densities, compositions, potential energies, pressure components, stresses, and diffusivities. In the current heterogeneous solid/liquid interface system, we confirm that the equilibrium conditions for the multiple in-plane equilibria among interfacial states can significantly deviate from the bulk phase diagram, consistent with the fact that GB complexion diagrams dramatically different from that of the bulk phases\cite{Cantwell14}. We found that the immiscible Cu and Pb (monotectic phase diagram) can form homogeneous monolayer alloy liquid states, which could coexist with the interfacial pre-freezing Pb solid and(or) the solid Cu step islands. The calculated data, especially the local pressure tensor components, could support the development of interfacial complexion (or phase) diagrams and offer clues to dig into the fundamental physics of the interfacial complexion (or phase), thus facilitating the advancement in interface (or the interfacial complexion) engineering.

\section{Methodology}
\label{sec:Method}

In this study, the many-body embedded atom method (EAM) potential for CuPb alloy parameterized by Hoyt et al.\cite{Hoyt03b} is employed to describe the interactions between atoms. It yields a reasonable prediction of the bulk binary phase diagram (including the melting points of pure Cu (1279K) and Pb (618K), as well as the mutual immiscibility between two metals in both liquid and solid phases). This interatomic potential has been used to explore the wetting\cite{Webb03} and spreading dynamics of Pb droplet on Cu substrates\cite{Heine05}, heterogeneous nucleations\cite{Palafox16} and equilibrium thermodynamic properties\cite{Palafox11} across the Cu(solid)/Pb(liquid) interfaces.

\begin{figure*}[!htb]
\centering
\includegraphics [width=0.95\textwidth] {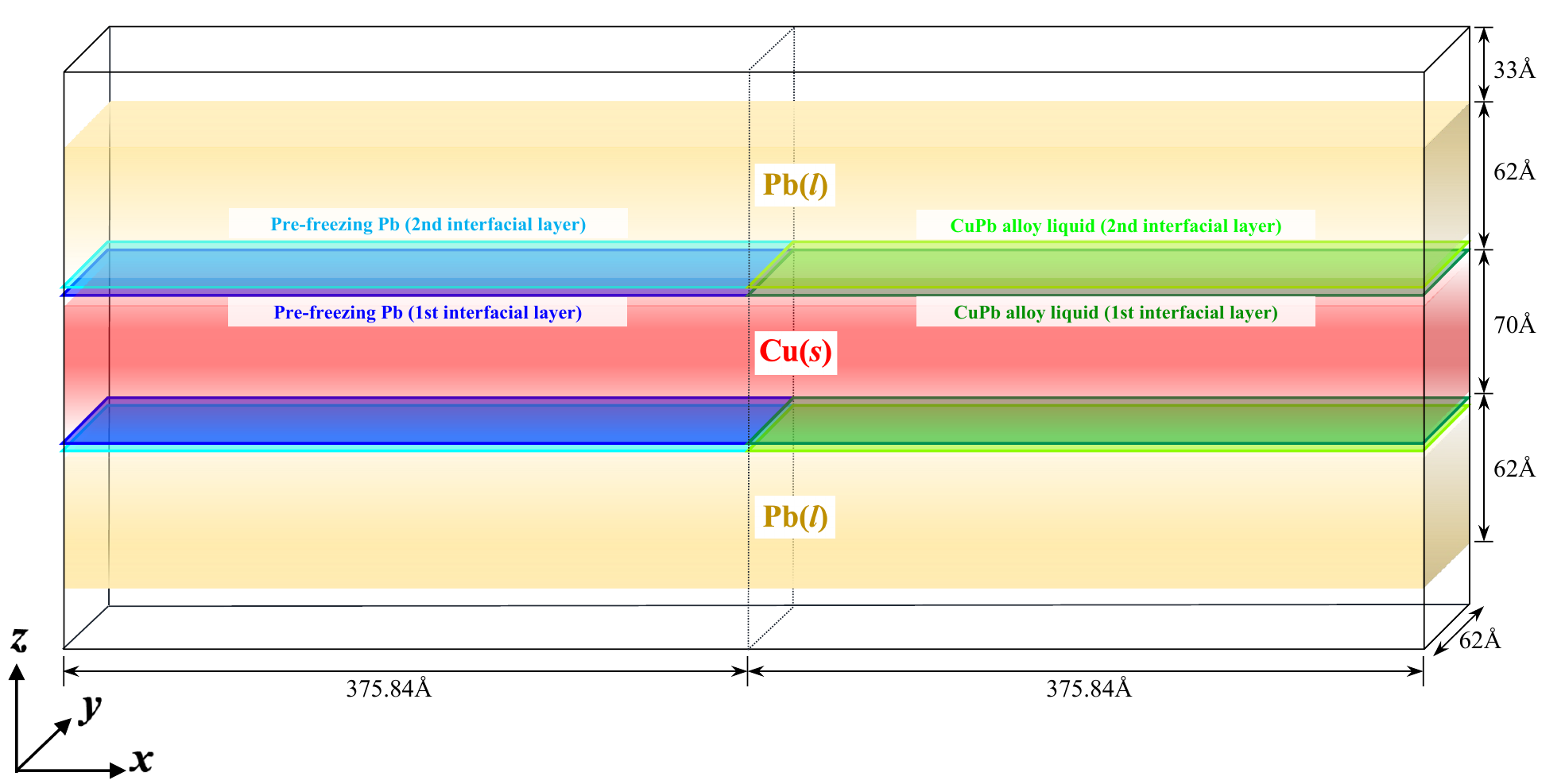}
\caption{The schematic diagram for the Cu(s)/Pb(l) interface system built with equilibrium MD simulations. Two Cu(s)/Pb(l) solid-liquid interfaces and two Pb(l)-vapor interfaces are parallel to the $xy$ plane. Besides the bulk solid, liquid, and vapor phases, the coexistences of multiple interfacial states are constructed by assembling two pre-equilibrated conglomerates half size of the whole simulation cell.}
\label{fig1}
\end{figure*}

The simulations are carried out using the large-scale atomic/molecular massively parallel simulator (LAMMPS)\cite{lammps} at constant $T=620$K, which is 2K above the freezing(or melting) point of Pb. The time-step and the No\'{s}e-Hoover thermostat relaxation time are set as 1.0fs and 0.1ps, respectively. The construction of the equilibrium Cu(111)/Pb(l) interfaces draws on the techniques described in Refs.\onlinecite{Palafox11,Yang12,Liang18}. The fcc(111)-oriented Cu crystal sample (84$\times$24$\times$11 unit cells, 133,056 Cu atoms) is constructed with (111) surface parallel to the $x$ and $y$ axes of the simulation cell using the lattice parameter at 620K (3.653266\AA) determined in a separate $NPT$ simulation. The cross-section ($xy$) dimension of the cell for simulating the crystal sample is 375.84\AA $\times 62.00$\AA, which is also employed in the separate preparation of two Pb liquid samples (45,000 Pb atoms in each sample) at the same temperature. The prepared crystal and two liquid samples are assembled by applying the periodic boundary conditions (PBC) along $x$ and $y$ to the initial conglomerate (around 223,000 atoms), and the crystal sample is set in between the two liquid samples. Along $z$ axis, the normal direction of the solid/liquid interface, we insert two vapor phases (33\AA \ thickness in $z$ each, see in Fig.\ref{fig1}) next to the two liquid samples, resulting in two liquid-vapor interfaces to aid in modeling the hydrostatic conditions both in bulk solid Cu and liquid Pb, in the following equilibration $NVT$ MD simulation. Note that the thickness of the solid and liquid samples is over 60\AA, which is thick enough to avoid the short-range interaction between solid/liquid interfaces and liquid-vapor interfaces. The linear momentum for the bulk Cu is fixed as zero to avoid its Brownian motion.

As based on the phase diagram (computationally verified by us using the semi-grand canonical Monte-Carlo\cite{Becker06}), the equilibrium mole fractions of Cu in bulk liquid Pb is around $X_\mathrm{Cu}^{l}(620$K$)=0.007(1)$, of Pb in bulk crystal Cu is negligible, or $X_\mathrm{Cu}^{s}=1$. So, prior to the equilibration $NVT$ MD simulation for the initial conglomerate, we randomly replace 310 Pb atoms in two bulk liquid sample parts into Cu atoms, yielding a composition value nearly equal to $X_\mathrm{Cu}^{l}(620$K$)$. Using this simple setup, i.e., put liquid Pb (with dilute solute atoms) on a clean Cu surface, two interfacial hexagonal pre-freezing Pb layers (rotated 6$^\circ$ relative to the underlying Cu(111) lattice) are easily formed within a few hundred picoseconds during the equilibration $NVT$ MD simulation, as reported in Refs.\onlinecite{Palafox11,Palafox16}. 

Until this work, the Cu(111)/Pb(l) interface at different temperatures had only been modeled in a ``clean'' fashion, i.e., two layers of pure Pb next to the complete Cu (111) surface\cite{Webb03,Heine05,Palafox11,Palafox16}. Enlightened by recent grand-canonically optimized atomistic simulation methodologies devoted to searching new GB complexion transitions with more possible sets of local structures, densities, and compositions\cite{Cantwell14,Cantwell20}. We examine in this work more possible equilibrium Cu(111)/Pb(l) interfacial states, which have different local composition values in the interfacial layers. Since the mass transportation of the solid/liquid interface is significantly greater than that of GBs, sufficient structural and compositional equilibration in both the interfacial and bulk regions can be easily achieved within several tens of nanoseconds via MD simulation, and there is no need to carry out additional grand-canonical Monte-Carlo simulation. Based on the pre-freezing Cu(111)/Pb(l) interface obtained as described in the previous paragraph, in the first pre-freezing Pb layer adjacent to the Cu surface, we randomly replace a portion of Pb atoms into Cu atoms, yielding a variety of in-plane Cu composition valves -- ranging from 0\% to 60\%, separated by a few percents. This operation is followed by an equilibration $NVT$ MD simulation lasting for 80 ns, in which both the microstructure and composition at the first two interfacial layers next to the Cu(111) surface are evolving and reaching new equilibrium interfacial states beyond the known interfacial pre-freezing phase.

As illustrated in the schematic Fig.\ref{fig1}, the final heterogeneous Cu(111)/Pb solid/liquid interface system that contains the coexistences of multiple interfacial states are constructed by assembling two pre-equilibrated conglomerates along $x$ axis and applying PBC in $x$ and $y$ axis of the combined simulation cell. One (of the two) conglomerates contains two interfacial pre-freezing Pb layers, and the second equilibrated conglomerate contains two interfacial alloy liquid layers. In this way, in each Cu(111)/Pb(l) interface, there are two parallel boundary lines between the coexisting pre-freezing Pb and interfacial CuPb alloy liquid states in the first two interfacial layers next to the Cu(111) surface, along $x$ axis. The dimension of the combined cell is $L_x \times L_y = 751.68 $\AA $\times 62.00$\AA, containing around 446,000 atoms. The two coexistence boundary lines are separated by over 370\AA \ in the distance, which is large enough to avoid the in-plane line-line interaction, which may cause the finite size effect of the boundary line free energy and their in-plane capillary fluctuations\cite{Liang18}. The $NVT$ simulation up to 30 ns ensures multiple interfacial states coexist, reaching structural and compositional equilibrium within the Cu(111)/Pb(l) interface.

After the combined simulation cell is equilibrated, a following-up 50 ns $NVT$ simulation is carried out to collect data. We divide the MD trajectories of this 50 ns long run into five sections for block-averaging. The simulation cell has two independent Cu(111)/Pb(l) interfaces, and each contains two interfacial layers interested. There are two parallel in-plane interfacial solid-liquid coexistence boundary lines within each interfacial layer interested. In total, we have twenty in-plane interfacial solid-liquid coexistence boundary lines samples (we randomly choose ten of them), and ten Cu(111)/Pb(l) interface samples for quantitative characterization and calculation.

The atomistic characterization methodologies developed for the chemically heterogeneous solid/liquid interfaces\cite{Palafox11,Yang12}, the interfacial layer containing the steps at chemically heterogeneous solid/liquid interfaces\cite{Liang18}, and the coexistence of mono-molecular layer confined crystal and its melt phase with coexistence boundary line\cite{Du18}, are employed to study various structural and thermodynamic properties for the equilibrium bulk Cu, bulk Pb, Cu(111)/Pb interface, interfacial CuPb alloy liquids, interfacial pre-freezing Pb, and the boundary lines between the interfacial solids and liquids.

The calculations include the fine- and coarse-scale profiles of various properties and their spatial variations along the $x$ axis for interfacial states and their in-plane transitional boundary line regions (or along the $z$ axis for bulk phases and their transitional regions). The structural and thermodynamic properties involved in this study include atomic number densities ($\rho_{\mathrm{Pb}}$ and $\rho_{\mathrm{Cu}}$), compositions ($X_{\mathrm{Pb}}$ and $X_{\mathrm{Cu}}$), potential energies ($Pe_{\mathrm{Pb}}$ and $Pe_{\mathrm{Cu}}$), pressure components ($p_{xx}$, $p_{yy}$ and $p_{zz}$), stresses (s), and diffusion coefficient components ($D_{xx}$, $D_{yy}$ and $D_{zz}$). 

In order to distinguish the quantities of the interfacial layers from the bulk phases, we add single-tilde ($^{\sim}$) and double-tilde ($^{\approx}$) over the symbols of various parameters to denote the first and the second interfacial layers next to the Cu(111) surface, respectively. We then extract different sets of fundamental parameters describing different phases and interfacial states from the uniform regions in the calculated profiles and employ superscript ``s'' and ``l'' to stand for (bulk or interfacial) solids and liquids, respectively.

\section{Results and Discussions}
\label{sec:Results}

\begin{figure*}[!htb]
\centering
\includegraphics [width=0.75\textwidth] {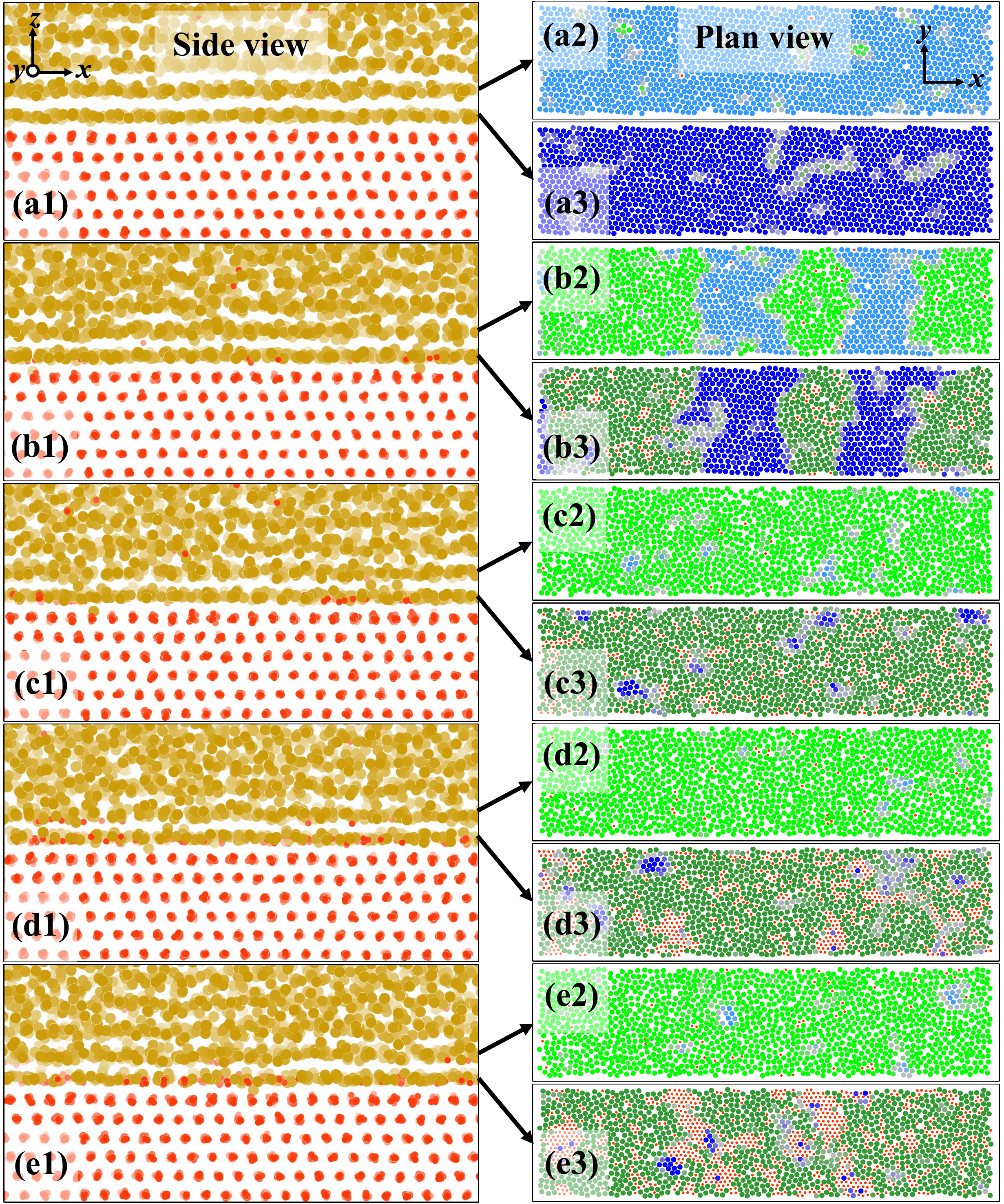}
\caption{MD snapshots of five pre-equilibrated Cu(111)/Pb(l) interfaces containing interfacial layers with different values of Cu compositions at 620K. A portion of the system is captured from the side view and plotted in the left-side panels. The direction along $x$, $y$, and $z$ axes are [1$\bar{2}$1], [10$\bar{1}$], and [111], respectively. Cu atoms are shown as smaller red circular plates in all panels, while Pb atoms are shown as brown circular plates with larger diameters in the left side panels. In the $xy$ plan views of the interfacial layers (in the right-side panels), Pb atoms are color-coded based on the values of ``bond''-angle orientational order parameter $q_6$\cite{Davidchack98}, blue for the interfacial solid state with higher values of $q_6$, green for the interfacial liquid state with lower values of $q_6$.}
\label{fig2}
\end{figure*}

Fig.\ref{fig2} shows the representative snapshots of five equilibrated single conglomerates containing interfacial layers with different values of Cu compositions. Fig.\ref{fig2}(a1-a3) demonstrate the obtained two stable interfacial pre-freezing Pb layers with nearly zero in-plane Cu compositions. We verify that i) the two pre-freezing Pb planes have a $\sim$1.9\% compressed lattice spacing compared to its bulk fcc lattice spacing, and ii) rotate 6$^\circ$ relative to the underlying Cu(111) lattice, identical to those reported by Palafox-Hernandez et al.\cite{Palafox11}. From Fig.\ref{fig2}(b1,b2,b3) to (e1,e2,e3), more equilibrated conglomerates containing interfacial layers with higher Cu composition values. Although the increasing numbers of Cu atoms from (a1) to (e1), each equilibrium bulk liquid phase in these cases has a relatively constant saturation solubility, i.e., $X_\mathrm{Cu}^{l}(620K)=0.007(1)$, suggesting that the varying excess Cu atoms in difference equilibrated interfacial systems are majorly concentrated within the first two interfacial layers. Numbers in parenthesis are the 95\% confidence level error-bars.

Fig.\ref{fig2}(b1-b3) demonstrate snapshots of the equilibrated interface from an initial setup with the Cu composition of the first Pb layer as 12\%. During the re-establishment of the equilibrium interfacial states, in-plane nucleation of the pre-freezing Pb island from the disordered binary alloy layer is observed. The pre-freezing islands grow laterally via a similar fashion of step growth of faceted silicon crystal from SiAl alloy melts\cite{Saidi17} until the Cu compositions in the remainder disordered interfacial CuPb mixture reach saturation values, i.e., $\sim20\%$ at the first layer adjacent to the Cu(111) surface and $\sim1.5\%$ at the second layer. Noticeable boundary lines between the interfacial pre-freezing Pb and interfacial CuPb alloy are seen. It is also observed here that the coupled parallel capillary waves of the neighboring boundary lines, probably bounding with the short-range disjoining potential among them\cite{Liang19}, are beyond the scope of the current study.

In the following panels of Fig.\ref{fig2}, panels (c1,c2,c3) to (e1,e2,e3), as the Cu composition in the interfacial layer increases, the nucleation and growth of the pre-freezing Pb island become rare. For example, for the conglomerates in which Cu composition $\sim$20\% in the first layer next to the Cu surface, Fig.\ref{fig2}(c3), the interfacial layers evolve into uniform Pb-rich solutions with individual Cu atoms and dynamic Cu clusters (mostly less than ten atoms) suspended within. In Fig.\ref{fig2}(d3),(e3), small solid Cu islands (with the size of several tens atoms) within the first interfacial layer (of the disordered CuPb alloy) are identified in the conglomerates with interfacial Cu composition approaching $30\%$ and higher. These localized Cu islands inherit the crystal lattice of the underlying Cu(111) surface and are relatively immobile compared with the suspending small Cu clusters not attached to the crystalline substrate. However, it is noticed that, in all the values of the interfacial Cu composition investigated in this work, the solid Cu islands are only found single-layer-wise, unlike the bi-layer pre-freezing Pb. This could be ascribed to the significant gradient in the Cu composition as the two interfacial layers are traversed from the crystal Cu side to the bulk liquid Pb side, i.e., the in-plane Cu composition decreased over one order of magnitude from the first interfacial layer to the second interfacial layer.

From the above-selected snapshots, one can observe that the interfacial disordered alloy states coexist well with the interfacial solids (i.e., pre-freezing Pb or solid Cu island). In order to further confirm the liquid nature of the disordered interfacial alloy states and study their fluidity as the function of the Cu composition. We examine the in-plane mean diffusion coefficients and the in-plane mean bond angle orientational order parameters\cite{Davidchack98} for both interfacial layers, i.e., $\langle {\tilde{D}_{xy}}\rangle$, $\langle \tilde{\tilde{D}}_{xy}\rangle$, $\langle \tilde{q_6}\rangle$, and $\langle \tilde{\tilde{q_6}}\rangle$, versus the equilibrium value of the Cu composition $\langle {\tilde{X}_\mathrm{Cu}}\rangle$ of the first interfacial layer adjacent to the Cu(111) surface. Over the whole composition range investigated, all diffusion coefficients of the first interfacial layer are greater than zero, although these interfacial data range in between 1/10 and 1/3 of the magnitude of $D$ for bulk liquid Pb phase at 620K (1.540(2)$\times10^{-5}$cm$^2$/s), they are at least four orders of magnitude greater than the diffusion coefficient in the bulk solid phase, indicating the disordered CuPb alloy states within the first interfacial layers have liquid-like dynamics. The maximum diffusivity and the minimum in-plane structural ordering of the first interfacial layer are found around the Cu composition value of 22\% -- corresponding to the uniform alloy liquid layer, which is free of pre-freezing Pb or solid Cu islands. The fluidity decease by either decreasing or increasing the Cu composition due to the rapidly developing of the pre-freezing Pb or the solid Cu island, respectively. The second interfacial layer shows more significant fluidity, which is reasonable as this layer is adjacent to the bulk Pb liquid.

Next, we present the equilibrium simulation results of the equilibrium heterogeneous Cu(111)/Pb(l) interface, which contains the coexistence of interfacial pre-freezing Pb domain and the CuPb alloy liquid domain(maximum fluidity, ${\tilde{X}_\mathrm{Cu}}\sim$22\%, $\tilde{\tilde{X}}_\mathrm{Cu}\sim$1.8\%,)  within the two interfacial layers. Fig.\ref{fig4}(a) demonstrate a full-size sideview snapshot of such equilibrium Cu(111)/Pb(l) interface from the viewpoint along [10$\bar{1}$]. In contrast to the rough type liquid Pb-vapor surface subjected to the capillary fluctuation, the Cu(111)/Pb(l) interface is flat, supporting a faceted nature for this heterogeneous solid/liquid interface at $T=620$K. Three local interfacial regions in boxes in Fig.\ref{fig4}(a) are zoomed in panels (b1,c1,d1), respectively, together with the corresponding fine-grained Cu and Pb density profiles along $z$, $\rho_\mathrm{Cu}(z)$ and $\rho_\mathrm{Pb}(z)$, in panels (b2,c2,d2). In (b1) and (b2), the Cu and Pb density profiles are resemble as previously reported data at $T=625$K\cite{Palafox11}. No interfacial alloying or no overlap of two density profiles at this region of the Cu(111)/Pb(l) interface is seen.

In the zoomed interfacial region shown in Fig.\ref{fig4}(c1) and (c2), the first liquid layer has more mixing of the Cu and Pb atoms, corresponding to the visible compositional overlap of the Cu and Pb density peaks at the first interfacial layer. Note, it has been reported at the Cu(100)/Pb(l) interface at 625K, the Cu and Pb density peaks also overlap, but in the surface Cu solid layer\cite{Palafox11}, rather than in the first interfacial Pb-rich alloy liquid layer, as zoomed in the region of current constructed Cu(111)/Pb(l) interface. In addition, the interface exhibits a less pronounced layer ordering structure (lower amplitude of the density peaks) in $\rho_\mathrm{Pb}(z)$ extending into the bulk liquid than is seen in panel (b1).

In panels (d1) and (d2) of Fig.\ref{fig4}, the zoomed interfacial region contains two coexisting interfacial states, i.e., the pre-freezing Pb and the CuPb alloy liquid states. $\rho_\mathrm{Cu}(z)$ profile barely changes from (b1) or (c1) to (d1). Nevertheless, there are minor differences related to composition or the oscillation amplitudes of the density profiles for $z>0$ due to the counting in the mixing of the two coexisting interfacial states.

In-plane snapshots of the two interfacial layers extracted from the zoomed region in Fig.\ref{fig4}(d1) are shown in Fig.\ref{fig5}(a) and (c), along with the contour plots of the 2D time-averaged densities, as well as the corresponding 2D Fourier structure factors\cite{Yang12}, see in panel (b) and (d) of Fig.\ref{fig5} and the insets therein. 

With the aid of color coding using the calculated bond angle orientational order ($q_6$) for each atom, the rough-type boundary lines between the in-plane coexisting pre-freezing Pb and the CuPb alloy liquid in both two interfacial layers can be tracked. In the vicinity of the equilibrium boundary lines, the perpetual in-plane growth and melt of the pre-freezing Pb edge via frequent (ps timescale) attachment/detachment of small Pb clusters are observed. In principle, the limiting kinetics that governs the line fluctuations and the line free energies could be further determined by examining the wave vector dependences of the line fluctuation relaxation times and the line fluctuation amplitudes\cite{Liang18}. These properties will be reported elsewhere.

The 2D time-averaged density maps of the two interfacial layers demonstrate that interfacial liquids and solids are well-equilibrated and coexistences of the multiple interfacial states and the adjacent bulk phases are highly stable over the entire simulation length. The apparent peak matrix on the left sides of Fig.\ref{fig5}(b) and (d) indicates the bi-layer-wise pre-freezing Pb is a robust structure as reported by Palafox-Hernandez et al.\cite{Palafox11}, even though we simulate a sample of the Cu(111)/Pb(l) interface with the cross-sectional size nearly ten times bigger than the work by Palafox-Hernandez et al. 

On the right side of Fig.\ref{fig5}(b), the first interfacial layer, the modulation of the underlying solid on the structure of the liquid alloy density field is hardly observed. However, faint yet visible residual Bragg peaks (identical to Cu(111) surface lattice) show the weak influence of the underlying crystalline Cu(111) surface. Correspondingly, such modulation becomes negligible on the right side of Fig.\ref{fig5}(d). No detectable protruding island structures due to the Cu island steps attached to the underneath Cu(111) surface are seen in the liquid alloy sides of both interfacial layers. The in-plane transition extents from pre-freezing Pb to alloy liquids for the two interfacial layers are similar, approximately over ten Pb atom diameters. The following paragraphs will show additional atomistic characterization of the thermodynamics properties for distinguishing differences between two interfacial layers regarding the in-plane coexistences of interfacial liquid and interfacial solid.

The profiles of various thermodynamics and dynamics quantities along $x$ axis for both interfacial layers are shown in Fig.\ref{fig6}, panels (a1-g1) show results for the first interfacial layer next to the Cu(111) surface, panels (a2-g2) show results for the second interfacial layer. $x=0$ corresponds to the in-plane Gibbs Dividing Surfaces (GDS)\cite{Frolov15b,Liang18}, where the boundary line excess number of Pb atoms is zero. $x<0$ for interfacial pre-freezing Pb solid and $x>0$ for interfacial alloy liquid.

The equilibrium coarse-scaled in-plane density profiles shown in Fig.\ref{fig6}(a1),(a2) count the particle number densities within the volumes for the two interfacial layers, the thickness of the interfacial layers are defined as the distances between the adjacent minima of the two fine-grained Pb density peaks next to the Cu(111) surface, as shown in Fig.\ref{fig4}(d2). All the density values of these two interfacial layers vary between the density values of the bulk Cu solid and bulk Pb liquid at the same temperature $T=620$K. The equilibrium values of the ``bulk'' regions of the interfacial solids and liquids for the two layers are listed in Tab.\ref{tab1}. Interestingly, in the first interfacial layer, the interfacial alloy liquid has a greater density value than the coexisting pre-freezing Pb due to the mixing of $\sim$22\% copper atoms, which has a smaller atomistic volume. In the second interfacial layer, because the Cu composition drops one order of magnitude, pre-freezing Pb in this layer owns larger density values than that of the dilute CuPb alloy liquids in the same layer, which has a nearly identical density value of the bulk Pb liquid. The in-plane decays of the densities for both layers span over a length scale of roughly 2.6 nm, i.e., the values of 10-90 widths\cite{Zhang22} extracted from the density profiles, ${\tilde{\delta}}_\mathrm{10-90}=2.7(2)$nm and ${\tilde{\tilde {\delta}}}_\mathrm{10-90}=2.4(3)$nm in Tab.\ref{tab1}. In contrast to the 10-90 width ($\delta_\mathrm{10-90}=0.8(1)$nm) of the Cu(111)/Pb(l) interface at this temperature, the relatively broader extents of the interfacial density decays can be attributed to the noticeable in-plane capillary waves along the boundary lines.

The equilibrium composition and potential energy proﬁles are shown in Fig.\ref{fig6}(b1,b2) and (c1,c2), the extracted equilibrium solubilities of Cu in interfacial liquids are 22.0(6)\% (first layer) and 1.8(1)\% (second layer), both show significant increases (around 34 times and three times greater) compared with the tiny solubility value of Cu in bulk Pb liquid phase (0.64(1)\%). Cu atoms are nearly immiscible for both interfacial layers with pre-freezing Pb, i.e., negligible solid state miscibilities ($\sim$0.2\%) for both layers, suggesting that the interfacial solid (pre-freezing) states are pure rather than solid-solutions. Despite that the monotectic phase diagram for bulk CuPb alloy (at $T=620$K) says that pure Cu solid coexists with nearly pure Pb liquid, it is interesting to note that the interfacial ``complexion (or phase) diagrams'' for the Cu(111)/Pb(l) interfacial layers bear a resemblance to that of the eutectic binary alloy systems, e.g., high liquid state miscibility and negligible solid state miscibility. The potential energy profiles (panels (c1) and (c2)) depict that atoms in the first interfacial layer have shallower potential minima than in the second. Within each layer, each atom in the solid state has a deeper potential minimum than in the alloy liquid state. The detailed potential energies of the Cu and Pb atoms belonging to both interfacial states within the two interfacial layers are listed in Tab.\ref{tab1}. 

The pressure components proﬁles in Fig.\ref{fig6}(d1-d2) present the in-plane coexistences of pre-freezing Pb and CuPb alloy liquid within the planar heterogeneous Cu(111)/Pb(l) interfacial layers, are under non-hydrodynamic conditions in which the pressure components are highly anisotropic and spatially non-homogeneous. All proﬁles in (d1) and (d2) show non-zero pressure, with absolute values on the magnitude of kbar at least three orders of magnitude greater than the hydrostatic pressure applied in bulk Cu solid or bulk Pb liquid phases. The $xx$ and $yy$ pressure components for the ``bulk'' regimes of the pre-freezing Pb (or CuPb alloy liquid) within both interfacial layers are equal within statistical uncertainties. Therefore only the averaged values of these two pressure components $p_{xy}=\frac{1}{2}(p_{xx}+p_{yy})$ are listed in Tab.\ref{tab1}. For CuPb alloy liquid sides, positive ${\tilde{p}_{xy}}=1.00(1)$kbar (first layer) and ${\tilde {\tilde {p}}}_{xy}=3.55(2)$kbar (second layer) are greater than those of their corresponding $p_{zz}$ components. For pre-freezing Pb sides, significant negative ${\tilde{p}_{xy}}=-6.30(4)$kbar (first layer) and significant positive ${\tilde {\tilde {p}}}_{xy}=9.42(5)$kbar (second layer) are found lower and much greater than those of their corresponding $zz$ components, respectively. As one travels across the interfacial coexistence boundary lines, from the pre-freezing Pb side to the alloy liquid side, ${\tilde{p}_{xy}}(x)$ (or ${\tilde {\tilde {p}}}_{xy}(x)$) show more remarkable relaxation in comparison with ${\tilde{p}_{zz}}(x)$ (or ${\tilde {\tilde {p}}}_{zz}(x)$), the $xx$ pressure components relax slightly before the $yy$ over pressure components in both interfacial layers, which give rise to the finite in-plane line lateral stresses shown in panels (f1) and (f2).

The panels (e1) and (e2) depict the detailed heterogeneity in the built-up non-hydrostatic mechanical stresses (defined as the difference between the longitudinal $zz$ and transverse $xy$ pressure components normal and parallel to the Cu(111)/Pb(l) interface) within interfacial layers. From the pre-freezing Pb side to the alloy liquid side, the mechanical stress decays from 3.69kbar to -3.16kbar in the first layer and increases from -10.08kbar to -2.99kbar in the second layer. The positive sign in the mechanical stress suggests that the first layer pre-freezing Pb is under lateral tension, whereas its in-plane coexisting alloy liquid is under lateral compression\cite{Lu22}. In the second interfacial layer, the coexisting pre-freezing Pb and the Pb-rich liquid states are both under lateral compression, although the former interfacial state withstands the highest magnitude of mechanical anisotropy. The in-plane lateral stress (defined as the difference between the pressure components normal $xx$ and parallel $yy$ to the in-plane states coexistence boundary line) profiles are plotted in panels (f1) and (f2) of Fig.\ref{fig6}. A single positive (or negative) peak in the later stress profile of the first (or the second) interfacial layer suggests the boundary line in the first (or the second) interfacial layer is in a state of stretching (or squeezing) along $y$ direction. The two peak structures in the two lateral stress profiles in (f1) and (f2) have close peak positions and relatively symmetric shapes. Therefore we deduce a strong short-range structural interaction may be binding the two boundary lines crossing the two interfacial layers, which deserves further investigations in the future.

\begin{figure*}[!htb]
\centering
\includegraphics [width=0.8\textwidth] {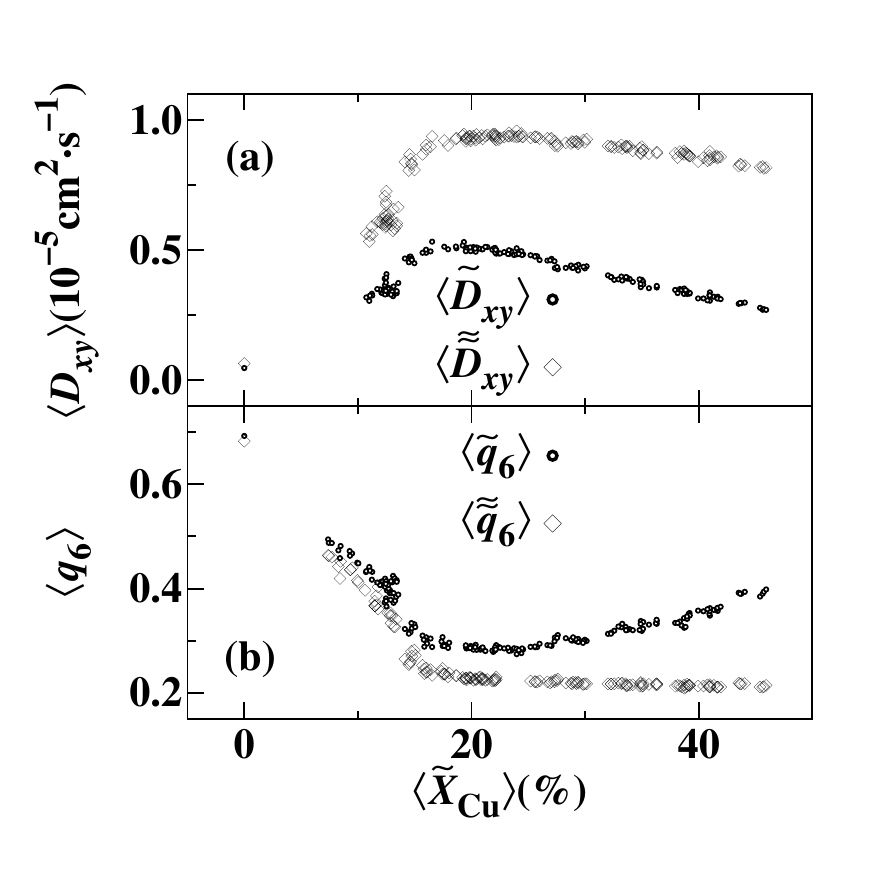}
\caption{Interfacial Cu composition $\langle {\tilde{X}_\mathrm{Cu}}\rangle$ dependences of (a) the in-plane mean diffusion coefficients, $\langle {\tilde{D}_{xy}}\rangle$, $\langle \tilde{\tilde{D}}_{xy}\rangle$, and (b) the in-plane mean bond angle orientational order parameters, $\langle \tilde{q_6}\rangle$, $\langle \tilde{\tilde{q_6}}\rangle$, for the first two interfacial layers adjacent to the Cu(111) surfaces at pre-equilibrated Cu(111)/Pb(l) interfaces at 620K.}
\label{fig3}
\end{figure*}

\begin{figure*}[!htb]
\centering
\includegraphics [width=0.8\textwidth] {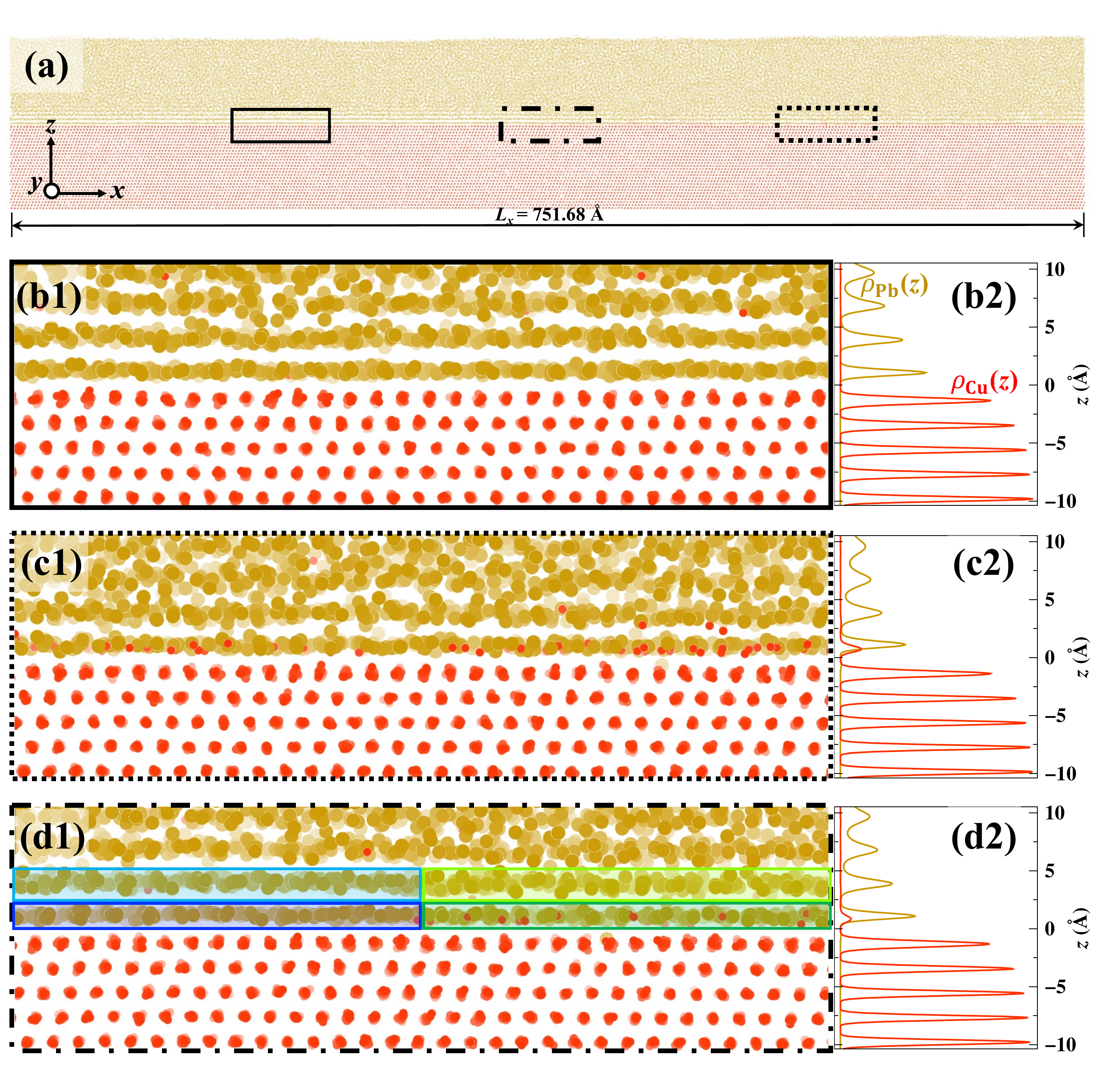}
\caption{(a) Full-size sideview snapshot of one equilibrium heterogeneous Cu(111)/Pb(l) interface at 620K, which contains the coexistence of interfacial pre-freezing Pb and the CuPb alloy liquids. Three local interfacial regions in boxes are zoomed in (b1), (c1), and (d1), respectively. (b2), (c2), and (d2) plot the corresponding fine-grained Cu (red) and Pb (brown) density proﬁles for the Cu(111)/Pb(l) interface along $z$ axis. $z<0$ for bulk Cu solid phase, $z>0$ for interfacial layers and bulk Pb liquid phase.}
\label{fig4}
\end{figure*}

\begin{figure*}[!htb]
\centering
\includegraphics [width=0.95\textwidth] {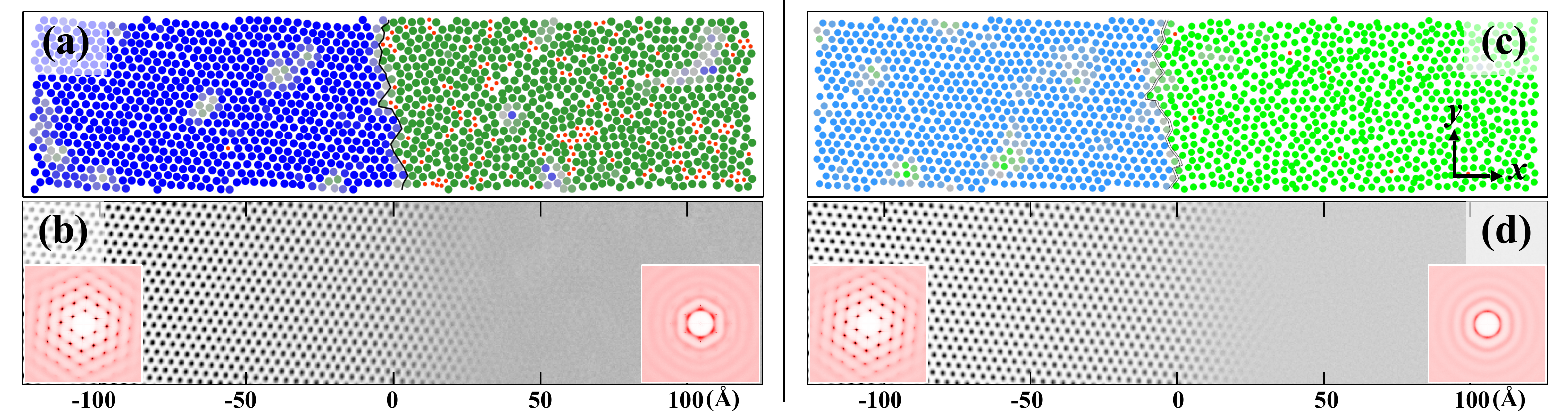}
\caption{(a) and (c) show instantaneous snapshots of the first and the second Cu(111)/Pb(l) interfacial layers extracted from the zoomed region in Fig.\ref{fig4}(d1), representing in-plane coexistence structure of interfacial solid states and interfacial liquid states. Cu atoms are represented with red color, Pb atoms are color-coded based on the values of $q_6$ order parameter, blue for interfacial solid states, green for interfacial liquid states. The boundary lines are plotted as black solid lines. (b) and (d) show the ``long exposure photography'' of the time-averaged 2D density contour maps, the 2D Fourier structure factors for the corresponding layers are plotted as insets in the two panels.}
\label{fig5}
\end{figure*}

\begin{figure*}[!htb]
\centering
\includegraphics [width=0.95\textwidth] {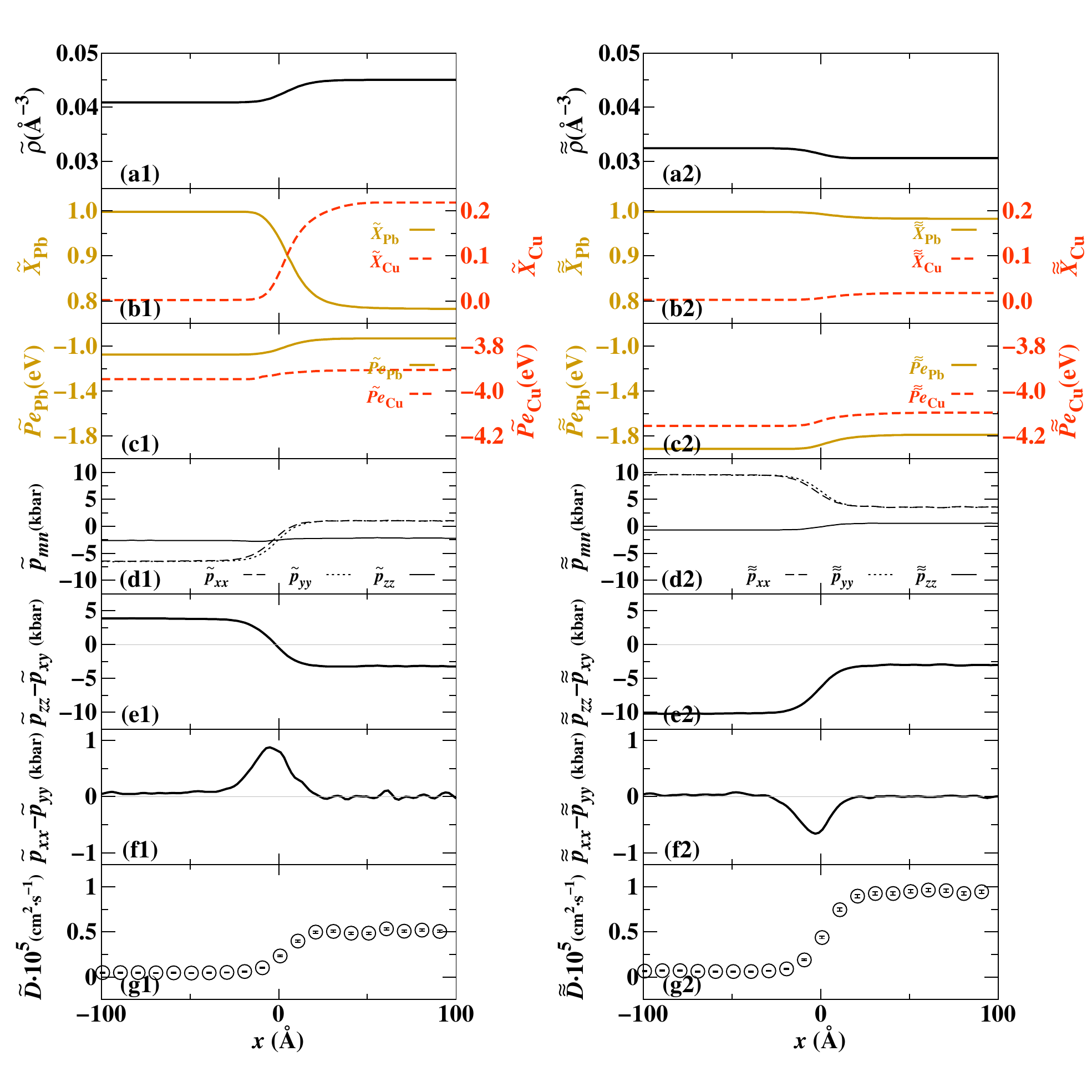}
\caption{In-plane coarse-scaled profiles of densities (a1.a2), compositions (b1,b2), potential energies (c1,c2), pressure components (d1,d2), interfacial stresses (e1,e2), lateral stresses (f1,f2), diffusion coefficients (g1,g2), for the two interfacial layers of the Cu(111)/Pb(l) interface at 620K. $x<0$ for interfacial pre-freezing Pb solids and $x>0$ for interfacial alloy liquids.}
\label{fig6}
\end{figure*}

To illustrate the spatial variation in the dynamic properties of the in-plane coexistence of interfacial solids and liquids, we compute the coarse-grained diffusion coefficients as the function of distances to the in-plane GDS, see in Fig.\ref{fig6}(g1),(g2). As the boundary lines are traversed from pre-freezing Pb sides to the alloy liquid sides, the diffusion coefficients start to increase from small values (0.047(2)$\times 10^{-5}$cm$^2$/s for the first interfacial layer, 0.064(2)$\times 10^{-5}$cm$^2$/s for the second interfacial layer) until they converge to their plateau ``bulk'' values, i.e., 0.509(5)$\times 10^{-5}$cm$^2$/s and 0.947(7)$\times 10^{-5}$cm$^2$/s, respectively, which are identical to the maximum diffusivity values reported in Fig.\ref{fig3}(a). Besides, we identify that the pre-freezing Pb domains in both layers are undergoing a random migration along $y$, similarly as a semi-2D ``crystal'' flake undergoes Brownian motion in liquid, thus explaining their finite value of the diffusion coefficients.

\begin{figure*}[!htb]
\centering
\includegraphics [width=0.8\textwidth] {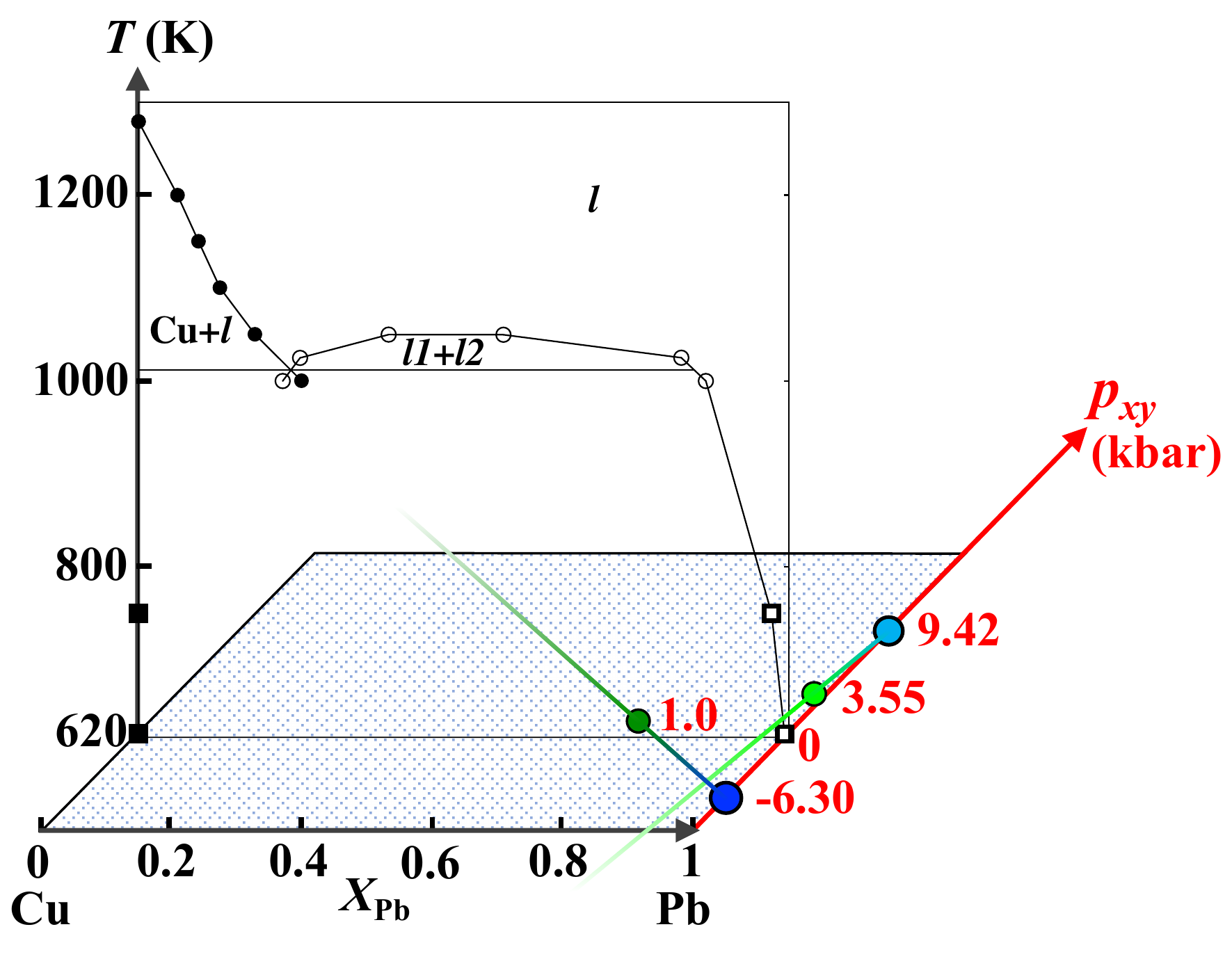}
\caption{Phase diagram of CuPb binary alloy system.
Filled and open circles are determined by a semi-Grand-Canonical MC method reported by Hoyt et al.\cite{Hoyt03b}, Filled and open squares are determined from the equilibrium MD simulations by Laird et al.\cite{Palafox16} at $T$=750K, and by current study at $T$=620K. One additional axis ($p_{xy}$) is added into the parameter space of the diagram for comparing the interfacial phase behavior with the bulk phase diagram. The dark and light blue circles represent the interfacial pre-freezing Pb states in the first and the second interfacial layers, respectively. The dark and light green circles represent the interfacial alloy liquid states in the first and the second interfacial layers, respectively.}
\label{fig7}
\end{figure*}

Based on the above characterization and computation of the multiple interfacial state coexistences, we obtain information that offers clues to further elucidate the relationship between the bulk phase diagram and the interfacial complexion (or phase) diagram. Being inspired by many previous works on developing GB diagrams, by overlaying the complexion diagram on the bulk phase diagram for a better illustration of the stability regions from a complete perspective\cite{Shi11,Cantwell14}. Here, instead of overlaying the CuPb bulk alloy phase diagram, we try to add one additional axis (i.e., transverse pressure components parallel to the Cu(111)/Pb(l) interface, $p_{xy}$) into the parameter space in Fig.\ref{fig7} for comparing the interfacial phase behavior with the bulk system. The employment of $p_{xy}$ is based on the following considerations. i) The values of $p_{xy}$ represent a majority of the local mechanical condition for the coexisting interfacial states, closely relating to their equations of state, chemical potentials (or the Gibbs free energies). A better option may be determining the local (including the interactions from adjacent materials) configurational integration of the partition function, which is exceedingly difficult. ii) There are experimental\cite{Emuna16,Hudon04} and theoretical\cite{Ma21} evidence shows substantial pressure-induced variation in the mutual miscibilities of heterogeneous binary alloy systems. The two schematic color $X_\mathrm{Pb}$--$p_{xy}$ lines for the two interfacial layers, respectively go through the interfacial ``phase'' equilibria compositions and local pressure components, labeled as the filled blue (pre-freezing Pb) and green (alloy liquid) circles. One could expect a  ``liquidus'' lines (containing more in-plane coexistence composition-local pressure data) which in equilibria with pre-freezing Cu (the corresponding ``solidus'' lines) at $T$ higher than 620K (note that Palafox-Hernandez and Laird observed that pre-freezing layer is no longer present at 750K\cite{Palafox16}).

Finally, with the current simulation results, it is interesting to discuss the validity of the recent thermodynamic formalism of the phase equilibria and phase rules into interfacial systems by Frolov and Mishin\cite{Frolov15b}. Based on their interfacial phase rule, the maximum possible number of interface phases $\nu_\mathrm{max}$ can coexist within an interface system composed of $\varphi=2$ bulk phases equal to $\nu_\mathrm{max}=k+3-\varphi$, in which $k$ is the number of chemical potentials $\mu_i$ for $k$ chemical components. For example, it predicts that an interface of a binary alloy system can support a maximum of three coexisting interface phases. Frolov and Mishin pointed out that their analysis was based on simplified isotropic treatment, which could not immediately apply to solid phases. Besides, the chemical potential of each chemical component was assumed as constant throughout the entire system, including bulk and interfacial phases. If we accept the idea that four interfacial states within the two interfacial layers coexist in Cu(111)/Pb(l) interface at 620K, as plotted in Fig.\ref{fig7}, we will find the predicted $\nu_\mathrm{max}=3$ contradicts the our current result.

However, if one considers the Cu and Pb atoms have distinct values of chemical potentials different from bulk values within different interfacial states and interfacial layers (note: this might be reasonable since the local symmetry broken structure for the interfacial layers should have unique configurational integrals contributing to their partition functions), for example, four more values of chemical potentials for interfacial Cu and Pb atoms, one would obtain a prediction of $\nu_\mathrm{max}=7$ maximum interfacial phases coexist in the Cu(111)/Pb(l) interface at 620K. In this way, the result of four coexisting interfacial states coexist in Cu(111)/Pb(l) interface in Fig.\ref{fig7} is in agreement with such phase rule prediction. The $\nu_\mathrm{max}=7$ prediction also implies more possible in-plane coexistences of interfacial states in Cu(111)/Pb(l) interface, which has not been substantially explored yet and deserves a independent future study.


\section{Summary}
\label{sec:summ}
Using the MD simulations, we uncover new interfacial states within the Cu(111)/Pb(l) at 620K. We predict the interfacial alloy liquid states dominate over the previously reported pre-freezing Pb solids with higher values of interfacial excess number of Cu atoms, and their potentially in-plane coexistences with the interfacial solid states (pre-freezing Pb or solid Cu step). The majority of previous discoveries of the multiple interfacial complexion (or phase) coexistence and transition are made in GBs and solid surface systems, rather than the heterogeneous solid/liquid interfacial systems. We provide here an atomistic simulation demonstration on the in-plane coexistences of multiple interfacial states within a planar heterogeneous Cu(111)/Pb(l) interface, including the interfacial pre-freezed metal Pb, the interfacial binary CuPb alloy liquids, and two boundary lines. Characterization of the proﬁles of a series of fundamental properties (e.g., densities, compositions, potential energies, pressure components, stresses, and diffusivities) for the in-plane equilibria of the interfacial solid and liquid states in the first two interfacial layers adjacent to the Cu(111) surface, are carefully carried out. Overall, the current construction of the constant temperature equilibrium Cu(111)/Pb(l) interface under ambient pressure conditions presents rich and in-plane varied mechanical, structural, and thermodynamical characteristics.

Four sets of the fundamental parameters for the four coexisting states are extracted from the two sets of the in-plane profiles (Fig.\ref{fig6}) of the two interfacial planes. These parameter sets suggest that each interfacial state examined has distinct values from each other and significantly deviates from those of the bulk solid and liquid phases. Before this study, there has been little discussion of interfacial complexion (or phase) diagrams in the solid/liquid interface systems. Our findings suggest that the interfacial “complexion (or phase) diagrams” for the Cu(111)/Pb(l) interface at a temperature just above the melting point of Pb bears a resemblance to that of the eutectic type binary alloy systems, in stead of the monotectic phase diagram for the bulk CuPb alloy. The current simulation results also enable us to evaluate the validity of the recent formalism of phase rules for the semi-2D interfacial systems. We propose that a more rigorous interface phase rule may need the consideration that atoms within different interfacial states and interfacial layers have distinct values of chemical potentials different from bulk values.

Preparation and tuning of the multi-interfacial state coexistences and transformations within the heterogeneous solid/liquid interfaces will facilitate the effective control of related processes with essential applications. For example, i) dynamics of wetting and spreading of molten metal on a heterogeneous solid metal substrate\cite{Yang17} can be strongly altered by the occurrence o f the interfacial pre-freezing\cite{Heine05} transitions. ii) The heterogeneous nucleation of Pb crystals on the Cu surface with pre-freezing Pb is found considerably more favorable than the Cu/Pb solid-liquid interface exhibits significant surface alloy\cite{Palafox16}. Our current simulation observation indicates a potential new method for quantitatively tuning the heterogeneous wetting and spreading of Pb droplet on the Cu(111) surface, as well as the heterogeneous nucleation of Pb crystal on the Cu(111) surface if one could effectively tune the allocation of the multiple-interfacial state coexistences and transformations within the Cu(111)/Pb(l).

\section{Acknowledgements}
This work is supported by the Chinese National Science Foundation (Grant No. 11874147), the Natural Science Foundation of Shanghai, the Natural Science Foundation of Chongqing (Grant No. cstc2021jcyj-msxmX1144).
\bibliographystyle{elsarticle-num}
\bibliography{ref}
\newpage

\newpage

\begin{sidewaystable}[h]
\caption{Summary of the thermodynamic parameters extracted from an equilibrium 620K Cu(111)/Pb(l) interface with in-plane coexistence structure of interfacial solid and interfacial liquid states.}
\begin{ruledtabular}
\begin{tabular}{lrlrlr}
\multicolumn{2}{c}{${\rm [Interfacial \ layer \ 1]} \ {\rm Pb}({\rm s}){\rm -alloy}({\rm l})$}&\multicolumn{2}{c}{${\rm [Interfacial \ layer \ 2]} \ {\rm Pb}({\rm s}){\rm -alloy}({\rm l})$}&\multicolumn{2}{c}{${\rm [Bulk \ interface]} \ {\rm Cu}(111){\rm /Pb}({\rm l})$}\\
\hline
${\tilde\rho}{\hspace{0.08em}}^{s{\rm Pb}}{\rm (nm^{-3})}$&40.90(1)&${\tilde{\tilde\rho}}{\hspace{0.08em}}^{s{\rm Pb}}{\rm (nm^{-3})}$&33.36(2)&$\rho{\hspace{0.08em}}^{s{\rm Cu}}{\rm (nm^{-3})}$&83.063(1)\\

${\tilde\rho}{\hspace{0.08em}}^{l{\rm alloy}}{\rm (nm^{-3})}$&45.03(8)&${\tilde{\tilde\rho}}{\hspace{0.08em}}^{l{\rm alloy}}{\rm (nm^{-3})}$&31.51(5)&$\rho{\hspace{0.08em}}^{l{\rm Pb}}{\rm (nm^{-3})}$&31.406(7)\\

$\tilde{X}{\hspace{0.08em}}^{s{\rm Pb}}_{\rm Cu}(\%)$&0.2(1)&${\tilde{\tilde{X}}}{\hspace{0.08em}}^{s{\rm Pb}}_{\rm Cu}(\%)$&0.2(1)&${X}{\hspace{0.08em}}^{s{\rm Cu}}_{\rm Cu}(\%)$&100\\

$\tilde{X}{\hspace{0.08em}}^{l{\rm alloy}}_{\rm Cu}(\%)$&22.0(6)&${\tilde{\tilde{X}}}{\hspace{0.08em}}^{l{\rm alloy}}_{\rm Cu}(\%)$&1.8(1)&${X}{\hspace{0.08em}}^{l{\rm Pb}}_{\rm Cu}(\%)$&0.64(1)\\

${\tilde {Pe}}{\hspace{0.08em}}^{s{\rm Pb}}_{\rm Cu}{\rm (eV)}$&-3.951(2)&${\tilde{\tilde {Pe}}}{\hspace{0.08em}}^{s{\rm Pb}}_{\rm Cu}{\rm (eV)}$&-4.156(1)&$ {Pe}{\hspace{0.08em}}^{s{\rm Cu}}_{\rm Cu}{\rm (eV)}$&-3.4547(1)\\
${\tilde {Pe}}{\hspace{0.08em}}^{l{\rm alloy}}_{\rm Cu}{\rm (eV)}$&-3.905(2)&${\tilde{\tilde {Pe}}}{\hspace{0.08em}}^{l{\rm alloy}}_{\rm Cu}{\rm (eV)}$&-4.094(4)&$ {Pe}{\hspace{0.08em}}^{l{\rm Pb}}_{\rm Cu}{\rm (eV)}$&-4.1033(2)\\
${\tilde {Pe}}{\hspace{0.08em}}^{s{\rm Pb}}_{\rm Pb}{\rm (eV)}$&-1.076(1)&${\tilde{\tilde {Pe}}}{\hspace{0.08em}}^{s{\rm Pb}}_{\rm Pb}{\rm (eV)}$&-1.914(1)&$ {Pe}{\hspace{0.08em}}^{s{\rm Cu}}_{\rm Pb}{\rm (eV)}$& N/A\\
${\tilde {Pe}}{\hspace{0.08em}}^{l{\rm alloy}}_{\rm Pb}{\rm (eV)}$&-0.933(6)&${\tilde{\tilde {Pe}}}{\hspace{0.08em}}^{l{\rm alloy}}_{\rm Pb}{\rm (eV)}$&-1.788(8)&$ {Pe}{\hspace{0.08em}}^{l{\rm Pb}}_{\rm Pb}{\rm (eV)}$&-1.8873(3)\\

${\tilde{p}}{\hspace{0.08em}}^{s{\rm Pb}}_{xy}{\rm (kbar)}$&-6.30(4)&${\tilde{\tilde{p}}}{\hspace{0.08em}}^{s{\rm Pb}}_{xy}{\rm (kbar)}$&9.42(5)&${p}{\hspace{0.08em}}^{s{\rm Cu}}_{xy}{\rm (kbar)}$&0.002(4)\\
${\tilde{p}}{\hspace{0.08em}}^{s{\rm Pb}}_{zz}{\rm (kbar)}$&-2.61(1)&${\tilde{\tilde{p}}}{\hspace{0.08em}}^{s{\rm Pb}}_{zz}{\rm (kbar)}$&-0.66(1)&${p}{\hspace{0.08em}}^{s{\rm Cu}}_{zz}{\rm (kbar)}$&0.001(1)\\
${\tilde{p}}{\hspace{0.08em}}^{l{\rm alloy}}_{xy}{\rm (kbar)}$&1.00(1)&${\tilde{\tilde{p}}}{\hspace{0.08em}}^{l{\rm alloy}}_{xy}{\rm (kbar)}$&3.55(2)&${p}{\hspace{0.08em}}^{l{\rm Pb}}_{xy}{\rm (kbar)}$&0.003(6)\\
${\tilde{p}}{\hspace{0.08em}}^{l{\rm alloy}}_{zz}{\rm (kbar)}$&-2.16(2)&${\tilde{\tilde{p}}}{\hspace{0.08em}}^{l{\rm alloy}}_{zz}{\rm (kbar)}$&0.56(2)&${p}{\hspace{0.08em}}^{l{\rm Pb}}_{zz}{\rm (kbar)}$&-0.004(4)\\

${\tilde{D}}{\hspace{0.08em}}^{s{\rm Pb}}_{xy}\cdot10^{5}(\rm {cm}^2\cdot{s}^{-1})$&0.047(2)&${\tilde{\tilde{D}}}{\hspace{0.08em}}^{s{\rm Pb}}_{xy}\cdot10^{5}(\rm {cm}^2\cdot{s}^{-1})$&0.064(2)&${D}{\hspace{0.08em}}^{s{\rm Cu}}\cdot10^{5}(\rm {cm}^2\cdot{s}^{-1})$&0.002(1)\\
${\tilde{D}}{\hspace{0.08em}}^{l{\rm alloy}}_{xy}\cdot10^{5}(\rm {cm}^2\cdot{s}^{-1})$&0.509(5)&${\tilde{\tilde{D}}}{\hspace{0.08em}}^{l{\rm alloy}}_{xy}\cdot10^{5}(\rm {cm}^2\cdot{s}^{-1})$&0.947(7)&${D}{\hspace{0.08em}}^{l{\rm Pb}}\cdot10^{5}(\rm {cm}^2\cdot{s}^{-1})$&1.540(2)\\
\end{tabular}
\end{ruledtabular}
\label{tab1} 
\end{sidewaystable}

\end{document}